\documentclass[pra,aps,showpacs,twocolumn]{revtex4-1}
\usepackage{amsmath}
\usepackage{amssymb}
\usepackage{graphicx}
\newcommand{\ket}[1]{|#1\rangle}
\newcommand{\bra}[1]{\langle #1|}

\newcommand{\abs}[1]{\lvert #1\rvert}
\begin{document}
\title{Ordering states with coherence measures}
\author{C. L. Liu}
\affiliation{Department of Physics, Shandong University, Jinan 250100, China}
\author{Xiao-Dong Yu}
\affiliation{Department of Physics, Shandong University, Jinan 250100, China}
\author{G. F. Xu}
\affiliation{Department of Physics, Shandong University, Jinan 250100, China}
\author{D. M. Tong}
\email{tdm@sdu.edu.cn}
\affiliation{Department of Physics, Shandong University, Jinan 250100, China}
\date{\today}
\begin{abstract}
 The quantification of quantum coherence has attracted a growing attention, and based on various physical contexts, several coherence measures have been put forward. An interesting question is whether these coherence measures give the same ordering when they are used to quantify the coherence of quantum states. In this paper, we consider the two well-known coherence measures, the $l_1$ norm of coherence and the relative entropy of coherence, to show that there are the states for which the two measures give a different ordering. Our analysis can be extended to other coherence measures, and as an illustration of the extension we further consider the formation of coherence to show that the $l_1$ norm of coherence and the formation of coherence, as well as the relative entropy of coherence and the coherence of formation, do not give the same ordering too.
\end{abstract}
\maketitle

\section{Introduction}
\label{intro}
Quantum coherence is considered to be one of the most important resources in quantum physics. It has many significant applications in various subjects, such as quantum optics \cite{Scully}, quantum information \cite{Nielsen}, quantum biology \cite{Huelga,Lambert}, and thermodynamics \cite{Aberg,Lostaglio,Rodriguez-Rosario}.
However, a general criterion to quantify the coherence of quantum states in information theoretic terms has only been proposed recently \cite{Baumgratz}, although the topic has ever been attempted in early papers \cite{Aberg1,Toloui,Marvian,Levi,Monras,Vogel,Smyth}. The general criterion has triggered the community's great interest \cite{Girolami,Bromley,Streltsov,Shao,Pires,Singh1,Du,Yuan,Yao,Mani,Cheng,Bera,Du1,Xi,Winter},and based on the criterion, various concrete measures of coherence have been given. The $l_1$ norm of coherence and the relative entropy of coherence, which satisfy the general criterion of quantifying the coherence of quantum states, were first suggested as two coherence measures based on distance in Ref.\cite{Baumgratz}. The coherence measure based on skew information \cite{Girolami}, and the coherence measure based on entanglement \cite{Streltsov} were then proposed. More expressions were subsequently shown to fulfill the general criterion of quantifying coherence, such as the coherence of formation, the distillable coherence, and the coherence cost \cite{Yuan,Winter}.

There have been several different expressions of coherence measures, as mentioned above. However, these different expressions, which are based on various physical contexts, result in different values of coherence for a state in general. In this situation, an interesting question is whether these coherence measures give the same ordering of states. That is, for two coherence measures $\mathcal {C}_A$ and $\mathcal {C}_B$, does $\mathcal {C}_A(\rho_1)\leq \mathcal {C}_A(\rho_2)$ imply  $\mathcal {C}_B(\rho_1)\leq \mathcal {C}_B(\rho_2)$ for any two states $\rho_1$ and $\rho_2$?  As two effective measures of coherence, one may expect $\mathcal {C}_A$ and $\mathcal {C}_B$ to give the same ordering for all the states even if numerical values of them are not quantitatively equal. In the paper, we address this issue. Our discussions focus on the two well-known coherence measures, the $l_1$ norm of coherence and the relative entropy of coherence, which have been widely used as a tool to investigate various aspects of coherence, such as the freezing phenomena of quantum coherence \cite{Bromley}, the complementarity relations for coherence measures \cite{Singh1,Cheng,Bera}, and the cohering and decohering power of quantum channels \cite{Mani}.  We will show that there are the states for which the two measures do not give the same ordering. As an extension of our analyses, we will also briefly discuss the other coherence measures, the $l_1$ norm of coherence and the coherence of formation as well as the relative entropy of coherence and the coherence of formation, showing that they do not give the same ordering too.

The paper is organized as follows. In Sec. 2, we review some notions related  to quantifying coherence and recall the three coherence measures under considered, i.e., the $l_1$ norm of coherence, the relative entropy of coherence, and the coherence of formation. In Sec. 3, we present our main results, i.e., we show that there are the states for which the $l_1$ norm of coherence and the relative entropy of coherence do not give the same ordering, and also give examples to show that the $l_1$ norm of coherence and the coherence of formation, as well as the relative entropy of coherence and the coherence of formation, do not give the same ordering too. Sec. 4 is our conclusions.

\section{The quantification of coherence}
\label{sec:1}
We review some notions related to quantifying coherence, such as incoherent states and incoherent operations, and recall the three coherence measures, the $l_1$ norm of coherence, the relative entropy of coherence, and the coherence of formation.

Let $\mathcal {H}$ be the Hilbert space for a $d$-dimensional quantum system. A particular basis of $\mathcal {H}$ is denoted as $\{\ket{i}, ~i=1,2,\cdot\cdot\cdot,d\}$, which is chosen according to the physical problem under discussion. Coherence of a state is then measured based on the basis chosen \cite{Baumgratz}. A state is called an incoherent state if and only if its density operator is diagonal in the basis, and the set of all the incoherent states is usually denoted as $\mathcal {I}$. Therefore, a density operator $\delta\in \mathcal {I}$ is of the form
$\delta=\sum^d_{i=1}\delta_{i}\ket{i}\bra{i}$.
All other states, which cannot be written as diagonal matrices in the basis, are called coherent states. Hereafter, we use $\rho$ to represent a general state, a coherent state or an incoherent state, and use $\delta$ specially to denote an incoherent state.

A completely positive trace preserving map, $\Phi(\rho)=\sum_iK_i \rho K_i^\dag$, is said to be an incoherent completely positive trace preserving (ICPTP) map, or an incoherent operation, if the Kraus operators $K_i$ satisfy not only $\sum_iK_i^\dag K_i=1$  but also $ K_i \mathcal {I} K_i^\dag \subseteq \mathcal {I} $, i.e., each $K_i$ maps an incoherent state to an incoherent state.

A functional $\mathcal {C}$ can be taken as a measure of coherence if it satisfies the four postulates \cite{Baumgratz}:
\\(C1) $\mathcal {C}(\rho)\geq0$, and $\mathcal {C}(\rho)=0$  if and only if $\rho\in \mathcal {I} $;\\
 (C2) $\mathcal {C}(\rho)\geq\mathcal {C}(\Phi_{ICPTP}(\rho))$ for all  ICPTP maps $\Phi_{ICPTP}$;\\
 (C3) $\mathcal {C}(\rho)\geq\sum_i \mathrm{Tr}(K_i\rho K_i^\dag)\mathcal {C}\left(K_i\rho K_i^\dag/\mathrm{Tr}(K_i\rho K_i^\dag)\right)$ for all $\{{K_i}\}$ with $\sum_n K^\dag_i K_i=\mathrm{I}$ and $ K_i \mathcal {I} K_i^\dag \subseteq \mathcal {I} $;\\
 (C4) $\mathcal {C}(\sum_ip_i\rho_i)\leq\sum_ip_i\mathcal {C}(\rho_i)$  for any set of states $\{\rho_i\}$ and any $p_i\geq0$ with $\sum_ip_i=1$.\\
Postulate (C1) imposes coherence measures to be a non-negative functional. (C2) and (C3) show the monotonicity  of coherence measures under incoherent operations and that under selective incoherent operations. (C4) means that coherence measures are nonincreasing under mixing of states. It is obvious that (C3) and (C4) imply (C2). These four postulates comprise a general criterion of defining a coherence measure.

In accordance with the general criterion, several coherence measures have been put forward. Out of them, there are the $l_1$ norm of coherence, the relative entropy of coherence, and the coherence of formation, which are considered in this paper.

The $l_1$ norm of coherence is defined as
 \begin{eqnarray}
 \mathcal {C}_{l_1}(\rho)=\sum_{i\neq j}|\rho_{ij}|,\label{L}
 \end{eqnarray}
 where $\rho_{ij}$ are entries of $\rho$. The coherence measure defined by the $l_1$ norm is based on the minimal distances of $\rho$ to the set of incoherent states $\mathcal {I}$, $\mathcal {C}_\mathcal {D}(\rho)=\min_{\delta\in \mathcal {I}}\mathcal {D}(\rho,\delta)$ with $\mathcal {D}$ being the $l_1$ norm, $0\leq\mathcal {C}_{l_1}(\rho)\leq{d-1}$. The upper bound is attained for the maximally coherent state $\ket{\varphi^d_{max}}=\frac{1}{\sqrt{d}}\sum_{i=1}^d\ket i$ \cite{Baumgratz}.

The relative entropy of coherence is defined as
 \begin{eqnarray}
\mathcal {C}_r({\rho})=\min_{\sigma\in \mathcal {I}}S(\rho\|\sigma)=S(\rho_{diag})-S(\rho),\label{entropy}
\end{eqnarray}
where $S(\rho\|\sigma)=\mathrm{Tr}(\rho\log_2\rho-\rho\log_2\sigma)$ is the quantum relative entropy, $S(\rho)=-\mathrm{Tr}(\rho\log_2\rho)$ is the von Neumann entropy, and $\rho_{diag}=\sum_i\rho_{ii}\ket{i}\bra{i}$ is the diagonal part of $\rho$. The coherence measure defined by the relative entropy is based on the minimal distances of $\rho$ to $\mathcal {I}$,  $\mathcal {C}_\mathcal {D}(\rho)=\min_{\delta\in \mathcal {I}}\mathcal {D}(\rho,\delta)$ with $\mathcal {D}$ being the the relative entropy, $0\leq\mathcal {C}_r(\rho)\leq\log_2{d}$. The coherence measure expressed by Eq.(\ref{entropy}) can also be derived based on the notion of distillation process, i.e., from the process that distills the maximally coherent states $\ket{\varphi^d_{max}}$ from $\rho$ under incoherent operations. It is the maximal rate at which the maximally coherent states can be obtained from the given state \cite{Winter}.

The coherence of formation, is defined as
\begin{equation}
  \mathcal {C}_f(\rho)=\min_{\{p_i, \ket{\varphi_i}\}}\sum_i p_i S((\ket{\varphi_i}\bra{\varphi_i})_{diag}),\label{cost}
\end{equation}
where $\rho=\sum_ip_i\ket{\varphi_i}\bra{\varphi_i}$ is any decomposition of $\rho$ into pure states $\ket{\varphi_i}$ with $p_i\geq 0$. Expression (\ref{cost}), as the coherence of formation, was first given in Ref. \cite{Aberg1}, and was proved to be a coherence measure, i.e., satisfying the four postulates, in Ref. \cite{Yuan}. The coherence measure expressed by Eq.(\ref{cost}) can also be derived based on the notion of coherence cost, i.e., from the process that prepares the state $\rho$ by consuming maximally coherent states $\ket{\varphi^d_{max}}$ under incoherent operations. It is the minimal rate at which the maximally coherent states have to be consumed to prepare the given state under incoherent operations \cite{Winter}.

\section{Main results}
\label{sec:2}
\subsection{Ordering states with coherence measures}

Before proceeding further, we first specify the notion of ordering states with coherence measures. For a coherence measure $\mathcal {C}$, the coherence of a state, $\mathcal {C}(\rho)$, is always a non-negative number, and therefore according to the numbers, all the states can be ranked in numerical order. For two coherence measures $\mathcal {C}_A$ and $\mathcal {C}_B$, we say that they give the same ordering if the following relations are satisfied for all the states $\rho_1$ and $\rho_2$,
\begin{equation}
\mathcal {C}_A(\rho_1)\leq \mathcal {C}_A(\rho_2)\Leftrightarrow \mathcal {C}_B(\rho_1)\leq \mathcal {C}_B(\rho_2).\label{order}
\end{equation}
Otherwise, we say that they do not give the same ordering. Relation (\ref{order}) means that $\mathcal {C}_A(\rho_1)\leq \mathcal {C}_A(\rho_2)$ if and only if $\mathcal {C}_B(\rho_1)\leq \mathcal {C}_B(\rho_2)$  for all the states $\rho_1$ and $\rho_2$, or  equally, $\mathcal {C}_A(\rho_1)\geq \mathcal {C}_A(\rho_2)$ if and only if $\mathcal {C}_B(\rho_1)\geq \mathcal {C}_B(\rho_2)$. Similar notions of ordering states have been widely used in entanglement measures \cite{Eisert,Zyczkowski,Virmani,Zyczkowski1,Wei,Wei1,Miranowicz,Miranowicz1,Verstraete,Miranowicz2,Miranowicz3,Ziman,Kinoshita,Horodecki} and other quantum-correlation measures \cite{De,Luo,Yeo,Ali,Al-Qasimi,Girolami1,Girolami2,Lang,Okrasa,Modi}.

We will show that the $l_1$ norm of coherence and the relative entropy of coherence do not give the same ordering of states, i.e., there are the states for which the two measures give a different ordering. To this end, we need to find the states for which $\mathcal {C}_{l_1}(\rho_1)<\mathcal {C}_{l_1}(\rho_2)$ and $\mathcal {C}_r(\rho_1)>\mathcal {C}_r(\rho_2)$, or $\mathcal {C}_{l_1}(\rho_1)>\mathcal {C}_{l_1}(\rho_2)$ and $\mathcal {C}_r(\rho_1)<\mathcal {C}_r(\rho_2)$. Hereafter, we refer to the pair of states for which the two coherence measures give a different ordering as the ordering-different pair for convenience.

{\subsection{The ordering-different pairs in $2$-dimensional systems}}

We first consider $2$-dimensional quantum systems. The density operators of a $2$-dimensional system can be generally written as
\begin{equation}
  \rho(x,y,z)=  \frac12\left(
  \begin{array}{ccc}
    1+z & x-iy\\
    x+iy& 1-z\\
  \end{array}
\right),\label{g2rho}
\end{equation}
with $x^2+y^2+z^2\leq 1$. $\rho(x,y,z)$ can be further expressed as
$\rho(x,y,z) =U_\alpha\rho(t,z)U^\dag_\alpha$,
where
\begin{equation}
\rho(t,z)=\frac12\left( \begin{array}{ccc} 1+z &t\\ t& 1-z\\ \end{array}\right), \label{rho2}
\end{equation}
with $t=\sqrt{x^2+y^2}$ and $t^2+z^2\leq 1$, and $U_\alpha=\text{diag}\left\{1,e^{i\alpha}\right\}$ is a diagonal and unitary matrix with $\alpha=-\arctan{\frac yx}$.
It means that $\rho(x,y,z)$ and $\rho(t,z)$ can be transformed into each other by an incoherent operation, which further implies that they correspond to the same coherence value for all the coherence measures due to postulate (C2). Therefore, we only need to consider the family of density operators with the form $\rho(t,z)$.

By substituting Eq. (\ref{rho2}) into Eqs. (\ref{L}) and (\ref{entropy}), we obtain the $l_1$ norm of coherence,
\begin{eqnarray}
\mathcal {C}_{l_1}\left(\rho\right)=t,\label{Lt}
\end{eqnarray}
and the relative entropy of coherence,
\begin{equation}
\mathcal {C}_{r}\left(\rho\right)=H(\frac12-\frac z2)-H(\frac12-\frac{\sqrt{z^2+t^2}}{2}),\label{Crtz}
\end{equation}
where $H(x)=-x\log_2x-(1-x)\log_2(1-x)$ is the binary Shannon entropy function.
${C}_{l_1}\left(\rho\right)$ is only dependent on $t$, while ${C}_r\left(\rho\right)$  is dependent
on both $t$ and $z$. Since ${C}_r\left(\rho\right)$ is an even function of $z$, we can further restrict our discussion to the range $0\leq z\leq 1$ without loss of generality.

Specially, when $\rho$ is a pure state, there is $\mathcal {C}_{r}\left(\rho\right)=H(\frac12-\frac{\sqrt{1-t^2}}{2})$ due to $t^2+z^2=1$. In this case, $\mathcal {C}_{r}\left(\rho\right)$ is an increasing function of $t$ just like $\mathcal {C}_{l_1}\left(\rho\right)$, and therefore the $l_1$ norm of coherence and the relative entropy of coherence always give the same ordering for all the $2$-dimensional pure states. There is no ordering-different pair in this case.

To find the ordering-different pairs for which the $l_1$ norm of coherence and the relative entropy of coherence give a different ordering, we further analyze the expression of $\mathcal {C}_{r}(\rho)$ without the restriction of pure states.
From Eq. (\ref{Crtz}), we have $\frac{\partial \mathcal {C}_{r}(\rho)}{\partial t}=
\frac{r}{2\sqrt{t^2+z^2}}\log_2{\frac{1+\sqrt{z^2+t^2}}{1-\sqrt{z^2+t^2}}}\geq0$ and $\frac{\partial \mathcal {C}_{r}(\rho)}{\partial z}=
\frac 12 \log_2{\frac{1-z}{1+z}}+\frac{z}{2\sqrt{t^2+z^2}}\log_2{\frac{1+\sqrt{z^2+t^2}}{1-\sqrt{z^2+t^2}}}
\geq 0$ for $0\leq z \leq1$, which means $\mathcal {C}_{r}(\rho)$ is an increasing function of $t$ as well as $z$.

With these knowledge, we may now try to find a pair of states $\rho_1$ and $\rho_2$, such that the ordering of $\mathcal {C}_{l_1}(\rho_1)$  and $\mathcal {C}_{l_1}(\rho_2)$ are different from that of $\mathcal {C}_r(\rho_1)$ and $\mathcal {C}_r(\rho_2)$.
 For this, we let
  \begin{eqnarray}
  \begin{aligned}
\rho_1=&\frac12\left(
  \begin{array}{ccc}
    1+z_1 & t_1\\
    t_1& 1-z_1\\
  \end{array}
\right),
~~
\rho_2=&\frac12\left(
  \begin{array}{ccc}
    1+z_2 & t_2\\
    t_2& 1-z_2\\
  \end{array}
\right).
\end{aligned}
\end{eqnarray}
Without loss of generality, we assume that $t_1 < t_2$, i.e., $\mathcal {C}_{l_1}(\rho_1)< \mathcal {C}_{l_1}(\rho_2)$. The question then turns to find $z_1$ and $z_2$ such that $\mathcal {C}_r(\rho_1)>\mathcal {C}_r(\rho_2)$, i.e.,
\begin{equation}
H(\frac12-\frac {z_1}{2})-H(\frac12-\frac{\sqrt{z_1^2+t_1^2}}{2})>H(\frac12-\frac {z_2}{2})-H(\frac12-\frac{\sqrt{z_2^2+t_2^2}}{2}).\label{inequality}
\end{equation}
For a given pair of  $t_1 < t_2$,  $z_1$ and $z_2$ satisfying Eq. (\ref{inequality}) may not exist in general. For instance, at $t_1=\frac35$ and $t_2=\frac45$, there do not exist $z_1$ and $z_2$ fulfilling Eq. (\ref{inequality}). Since $\mathcal {C}_{r}(\rho)$ is an increasing function of $t$ and $z$, the maximal value of the left side of Eq. (\ref{inequality}) is appearing at $z_1=\sqrt{1-t_1^2}$,  and the minimal value of the right side is appearing at $z_2=0$. Hence, we obtain the necessary and sufficient condition of $t_1$ and $t_2$,
\begin{equation}
 H(\frac{1-\sqrt{1-t_1^2}}{2}) > 1-H(\frac{1-t_2}{2}),\label{tcondition}
\end{equation}
under which $z_1$ and $z_2$ exist. It means that $z_1$ and $z_2$ satisfying the inequality (\ref{inequality}) exist if and only if $t_1$ and $t_2$ ($t_1\leq t_2$) satisfy the relation (\ref{tcondition}).

\begin{figure}[ht]
 \centering
 \includegraphics[width=84mm]{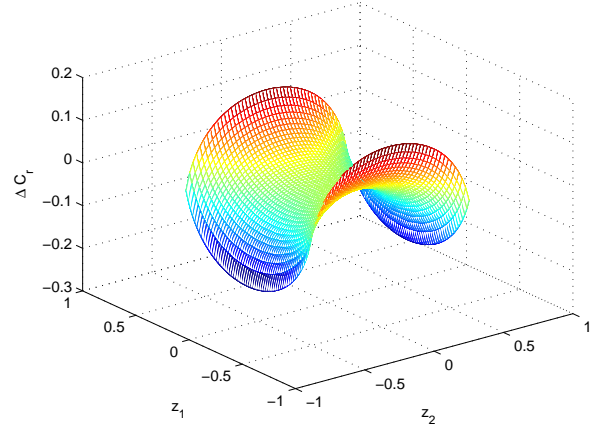}
 \caption{(Color online). An illustration to show that there are many solutions of $z_1$ and $z_2$ satisfying  $\Delta\mathcal {C}_r=\mathcal {C}_r(\rho_1)-\mathcal {C}_r(\rho_2)>0$. Here, we have taken $t_1=\frac45$ and $t_2=\frac2{\sqrt{6}}$. All the pairs of states $\rho_1$ and $\rho_2$ defined by $z_1$ and $z_2$ in the domain for $\Delta \mathcal{C}_r>0$ fulfill $\mathcal {C}_{l_1}(\rho_1) < \mathcal {C}_{l_1}(\rho_2)$ and $\mathcal {C}_r(\rho_1)>\mathcal {C}_r(\rho_2)$. }
 \label{fig:caitu}
\end{figure}
Therefore, to find $\rho_1$ and $\rho_2$ that satisfy both $\mathcal {C}_{l_1}(\rho_1) < \mathcal {C}_{l_1}(\rho_2)$ and $\mathcal {C}_r(\rho_1)>\mathcal {C}_r(\rho_2)$, one may first choose $t_1$ and $t_2$ ($t_1<t_2$) with the aid of Eq. (\ref{tcondition}) and then find $z_1$ and $z_2$ ($0\leq z_1, ~z_2\leq 1$) by using Eq. (\ref{inequality}).  For example, if we take $t_1=\frac45$ and $t_2=\frac2{\sqrt{6}}$, which satisfy Eq.(\ref{tcondition}), all the $z_1$ and $z_2$ in the domain with $\Delta\mathcal {C}_r=\mathcal {C}_r(\rho_1)-\mathcal {C}_r(\rho_2)>0$ are given in Fig. 1. Clearly, there are infinitely many pairs of states satisfying $\Delta\mathcal {C}_r>0$.
An explicit expression of $\rho_1$ and $\rho_2$ reads
\begin{equation}
\rho_1=\left(
  \begin{array}{ccc}
    \frac45 & \frac25 \\
    \frac25 & \frac15\\
  \end{array}
\right),~~ \rho_2=\left(
  \begin{array}{ccc}
   \frac12&\frac1{\sqrt{6}}\\
   \frac1{\sqrt{6}}&\frac12\\
  \end{array}
\right),
\label{ccl}
\end{equation}
for which, $\mathcal {C}_{l_1}(\rho_1)=\frac45< \mathcal {C}_{l_1}(\rho_2)=\frac2{\sqrt{6}}$ but $\mathcal {C}_r(\rho_1)=0.7219>\mathcal {C}_r(\rho_2)=0.5576$.\\

\subsection{The ordering-different pairs in high dimensional systems}

We now consider high dimensional quantum systems. The above discussion is only on $2$-dimensional states. It shows that there are the pairs of states for which
the $l_1$ norm and the relative entropy of coherence do not give the same ordering. This conclusion can be easily extended to the case of high dimensional density operators. In fact, any two $d (\ge 3)$-dimensional states defined by
\begin{equation}
\rho^{(d)}_1=\frac12( \delta_1^{(d-2)}\oplus\rho_1),~~~~ \rho_2^{(d)}=\frac12( \delta_2^{(d-2)}\oplus\rho_2),\label{rhod}
\end{equation}
where $\delta_1^{(d-2)}$ and $\delta_2^{(d-2)}$ are $(d-2)$-dimensional incoherent states, necessarily satisfy $\mathcal {C}_{l_1}(\rho_1^{(d)})<\mathcal {C}_{l_1}(\rho_2^{(d)})$ and $\mathcal {C}_r(\rho_1^{(d)})>\mathcal {C}_r(\rho_2^{(d)})$ as long as $\rho_1$ and $\rho_2$ do. This is because $\mathcal {C}_{l_1}(\rho_i^{(d)})=\frac12\mathcal {C}_{l_1}(\rho_i)$ and $\mathcal {C}_r(\rho_i^{(d)})=\frac12\mathcal {C}_r(\rho_i)$, $i=1,2$.

It is worth noting that although the $l_1$ norm of coherence and the relative entropy of coherence give the same ordering for all the $2$-dimensional pure states, it is not true for the $d (\geq3)$-dimensional case. There are pairs of high dimensional pure states for which the $l_1$ norm of coherence and the relative entropy of coherence do not give the same ordering. For example, in the $3$-dimensional case,
\begin{equation}
  \begin{split}
    \ket{\varphi^{(3)}_1}&=\sqrt{\frac{12}{25}}\ket{1}+\sqrt{\frac{12}{25}}\ket{2}+\sqrt{\frac1{25}}\ket{3},\\
    \ket{\varphi^{(3)}_2}&=\sqrt{\frac7{10}}\ket{1}+\sqrt{\frac15}\ket{2}+\sqrt{\frac1{10}}\ket{3},
  \end{split}
\label{psi12}
\end{equation}
is a pair of states for which the $l_1$ norm of coherence and the relative entropy of coherence do not give the same ordering, since $\mathcal {C}_{l_1}(\ket{\varphi_1})=1.5143<\mathcal {C}_{l_1}(\ket{\varphi_2})=1.5603$ but $\mathcal {C}_{r}(\ket{\varphi_1})=1.2023> \mathcal {C}_{r}(\ket{\varphi_2})=1.1568$.
With the aid of the $3$-dimensional case, we can find pairs of higher dimensional pure states for which the $l_1$ norm of coherence and the relative entropy of coherence do not give the same ordering. Indeed, any pairs of $d(\geq 4)$-dimensional pure states defined by
\begin{equation}
  \begin{split}
    \ket{\varphi^{(d)}_1}&=\alpha\ket{\varphi^{(3)}_1}+\sum_{i=4}^d\beta_i\ket{i},\\
    \ket{\varphi^{(d)}_2}&=\alpha\ket{\varphi^{(3)}_2}+\sum_{i=4}^d\beta_i\ket{i},
  \label{lyt}
    \end{split}
\end{equation}
with $\abs{\alpha}^2+\sum_{i=4}^d\abs{\beta_i}^2=1$ and $0<\abs{\alpha}<1$ necessarily satisfy $\mathcal{C}_{l_1}(\ket{\varphi^{(d)}_1})<\mathcal{C}_{l_1}(\ket{\varphi^{(d)}_2})$ and $\mathcal{C}_r(\ket{\varphi^{(d)}_1})>\mathcal{C}_r(\ket{\varphi^{(d)}_2})$ as long as $\ket{\varphi^{(3)}_1}$ and $\ket{\varphi^{(3)}_2}$ do (see the Appendix for the proof).

\subsection{Further discussions}

After having shown that the $l_1$ norm of coherence and the relative entropy of coherence do not give the same ordering of states, we now extend our discussion to one more coherence measure, the coherence of formation. We wonder whether the $l_1$ norm of coherence and the coherence of formation, as well as the relative entropy of coherence and the coherence of formation, give the same ordering of states. We will briefly discuss the issue in this subsection. In fact, with the aid of the above calculations, it is very easy to confirm that there are infinitely many ordering-different pairs for these coherence measures too.

We first address the $l_1$ norm of coherence and the coherence of formation. Form Eqs. (\ref{entropy}) and (\ref{cost}), we have $\mathcal{C}_f(\ket{\varphi})=\mathcal{C}_r(\ket{\varphi})$ for all pure states $\ket{\varphi}$, which implies that all the pure states that satisfy
$\mathcal {C}_{l_1}(\ket{\varphi_1})<\mathcal {C}_{l_1}(\ket{\varphi_2})$ and $\mathcal {C}_r(\ket{\varphi_1})>\mathcal {C}_r(\ket{\varphi_2})$ must fulfill $\mathcal {C}_{l_1}(\ket{\varphi_1})<\mathcal {C}_{l_1}(\ket{\varphi_2})$ and $\mathcal {C}_f(\ket{\varphi_1})>\mathcal {C}_f(\ket{\varphi_2})$. Therefore, all pairs of the $3$-dimensional states defined by Eq. (\ref{psi12}) and the $d(\geq 4)$-dimensional states defined by Eq. (\ref{lyt}) are the ordering-different pairs for the $l_1$ norm of coherence and the coherence of formation too. Note that there is no ordering-different pair in the $2$-dimensional case, since both $\mathcal {C}_{l_1}(\rho)=t$ and $\mathcal {C}_f(\rho)=H(\frac12+\frac{\sqrt{1-t^2}}{2}) $ \cite{Streltsov,Yuan} are increasing functions of $t$ and therefore they give the same ordering for all the $2$-dimensional states.

We now address the relative entropy of coherence and the coherence of formation. Since the coherence of formation and the $l_1$ norm of coherence give the same ordering for all the $2$-dimensional states as shown above, i.e., $\mathcal {C}_f(\rho_1)<\mathcal {C}_f(\rho_2)$ if and only if $\mathcal {C}_{l_1}(\rho_1)<\mathcal {C}_{l_1}(\rho_2)$, all the $2$-dimensional states that satisfy  $\mathcal {C}_{l_1}(\rho_1) < \mathcal {C}_{l_1}(\rho_2)$ and $\mathcal {C}_r(\rho_1)>\mathcal {C}_r(\rho_2)$ must fulfill $\mathcal {C}_f(\rho_1) < \mathcal {C}_f(\rho_2)$ and $\mathcal {C}_r(\rho_1)>\mathcal {C}_r(\rho_2)$. Therefore, all the $2$-dimensional ordering-different pairs for $\mathcal {C}_{l_1}$ and $\mathcal {C}_r$ are the ordering-different pairs for $\mathcal {C}_f$ and $\mathcal {C}_r$ too. As pointed out in the case of the $l_1$ norm of coherence and the relative entropy of coherence, there are infinitely many such pairs of states, and one may obtain them with the aid of Eqs. (\ref{inequality}) and (\ref{tcondition}). An explicit expression of $\rho_1$ and $\rho_2$ is given in Eq. (\ref{ccl}), for which $\mathcal {C}_f(\rho_1)=0.7219 < \mathcal {C}_f(\rho_2)=0.7440$ and $\mathcal {C}_r(\rho_1)=0.7219>\mathcal {C}_r(\rho_2)=0.5576$.

Besides, just like the result that any two entanglement measures coinciding on pure states are either identical or give a different ordering \cite{Virmani}, we can also obtain a similar result for coherence measures. Let us consider two coherence measures $\mathcal {C}_A$ and $\mathcal {C}_B$ that coincide on all pure states. Note that for any state $\rho$ and any sufficiently small $\epsilon>0$, there must exist pure states $\ket{\phi}$ and $\ket{\psi}$ such that $\mathcal {C}_A(\ket{\psi})-\epsilon = \mathcal {C}_A(\rho)= \mathcal {C}_A(\ket{\phi})+\epsilon$. If  $\mathcal {C}_A$ and $\mathcal {C}_B$ give the same ordering on all states,  we have $\mathcal {C}_B(\ket{\psi})\geq \mathcal {C}_B(\rho)\geq\mathcal {C}_B(\ket{\phi})$. As $\mathcal {C}_A$ and $\mathcal {C}_B$ coincide on pure states, we get $\mathcal {C}_A(\ket{\psi})\geq \mathcal {C}_B(\rho)\geq\mathcal {C}_A(\ket{\phi})$, which further leads to $\mathcal {C}_A(\rho)+\epsilon\geq\mathcal {C}_B(\rho)\geq\mathcal {C}_A(\rho)-\epsilon$. Taking $\epsilon$ to zero, we finally obtain $\mathcal {C}_A(\rho)=\mathcal {C}_B(\rho)$, which implies  that any two coherence measures that coincide on pure states are either identical or give a different ordering. This result can be used to determine whether some coherence measures, for instance  $\mathcal {C}_r$ and $\mathcal {C}_f$,  give the same ordering. However, it is not applicable to many other coherence measures, for example $\mathcal {C}_{l_1}$ and $\mathcal {C}_r$, which do not coincide on pure states.

\section{Conclusions}
The issue of ordering states with coherence measures is first discussed in this paper. We have shown that the two well-known coherence measures, the $l_1$ norm of coherence and the relative entropy of coherence, do not give the same ordering of states. There are infinitely many ordering-different pairs for which the $l_1$ norm of coherence and the relative entropy of coherence give a different ordering. Detailed calculations show that the ordering-different states include the $2$-dimensional mixed states, the $d (\geq 3)$-dimensional mixed states and the $d (\geq 3)$-dimensional pure states, but exclude 2-dimensional pure states since all the coherence measures give the same ordering for 2-dimensional pure states.

Our analysis can be extended to other coherence measures, and as an illustration of the extension we have shown that the $l_1$ norm of coherence and the formation of coherence, as well as the relative entropy of coherence and the coherence of formation, do not give the same ordering too. Our results indicate that at least each pair of the three coherence measures, the $l_1$ norm of coherence, the relative entropy of coherence and the coherence of formation, give a different ordering of states, although it remains an open question to examine whether all other coherence measures give a different ordering of states.

Noting that any two states in an ordering-different pair cannot be transformed into each other by incoherent operations, our result may be helpful in determining what kinds of states cannot be transformed into each other by incoherent operations.
\section{Appendix}

We show that any two $d(\geq 4)$-dimensional pure states defined by Eq. (\ref{lyt})
necessarily satisfy $\mathcal{C}_{l_1}(\ket{\varphi^{(d)}_1})<\mathcal{C}_{l_1}(\ket{\varphi^{(d)}_2})$ and $\mathcal{C}_r(\ket{\varphi^{(d)}_1})>\mathcal{C}_r(\ket{\varphi^{(d)}_2})$ as long as $\ket{\varphi_1}$ and $\ket{\varphi_2}$ do. To this end, we only need to prove the following theorem:  If the two $(d-1)$-dimensional states expressed as
 \begin{align}
 \ket{\varphi^{(d-1)}_1}=\sum_{i=1}^{d-1}a_i\ket{i},~~~~ \ket{\varphi^{(d-1)}_2}&=\sum_{i=1}^{d-1}b_i\ket{i},
  \label{eq:defphid-1}
\end{align}
satisfy $\mathcal{C}_{l_1}(\ket{\varphi^{(d-1)}_1})<\mathcal{C}_{l_1}(\ket{\varphi^{(d-1)}_2})$ and $\mathcal{C}_r(\ket{\varphi^{(d-1)}_1})>\mathcal{C}_r(\ket{\phi^{(d-1)}_2})$, then the two $d$-dimensional states defined by
\begin{align}
    \ket{\varphi^{(d)}_1}=\alpha_d\ket{\varphi^{(d-1)}_1}+\beta_d\ket{d},~~
    \ket{\varphi^{(d)}_2}=\alpha_d\ket{\varphi^{(d-1)}_1}+\beta_d\ket{d}
  \label{eq:defphid}
\end{align}
with $\abs{\alpha_d}^2+\abs{\beta_d}^2=1$ and $0<\abs{\alpha_d}<1$, satisfy  $\mathcal{C}_{l_1}(\ket{\varphi^{(d)}_1})<\mathcal{C}_{l_1}(\ket{\varphi^{(d)}_2})$ and $\mathcal{C}_r(\ket{\varphi^{(d)}_1})>\mathcal{C}_r(\ket{\varphi^{(d)}_2})$.

We now prove the theorem.

By the definition of the $l_1$-norm of coherence, we have
\begin{align}
  \begin{split}
    \mathcal{C}_{l_1}(\ket{\varphi^{(d-1)}_1})=&2\sum_{1\le i<j\le d-1}\abs{a_ia_j},
  \end{split}
  \label{eq:cl1phid-1}
\end{align}
and
\begin{align}
  \begin{split}
    \mathcal{C}_{l_1}(\ket{\varphi^{(d)}_1})=&2\abs{\alpha_d}^2\sum_{1\le i<j\le d-1}\abs{a_ia_j}+2\abs{\alpha_d\beta_d}\sum_{i=1}^{d-1}\abs{a_i},\\
    =&2\abs{\alpha_d}^2\sum_{1\le i<j\le d-1}\abs{a_ia_j}+2\abs{\alpha_d\beta_d}\sqrt{(\sum_{i=1}^{d-1}\abs{a_i})^2}\\
    =&2\abs{\alpha_d}^2\sum_{1\le i<j\le d-1}\abs{a_ia_j}\\
    &+2\abs{\alpha_d\beta_d}\sqrt{1+2\sum_{1\le i<j\le d-1}\abs{a_ia_j}},\\
    =&\abs{\alpha_d}^2C_{l_1}(\ket{\varphi^{(d-1)}_1})\\
    &+2\abs{\alpha_d\beta_d}\sqrt{1+C_{l_1}(\ket{\varphi^{(d-1)}_1})},
  \end{split}
  \label{eq:cl1phi1d}
\end{align}
and similarly,
\begin{align}
  \begin{split}
   \mathcal{C}_{l_1}(\ket{\varphi^{(d-1)}_2})=&2\sum_{1\le i<j\le d-1}\abs{b_ib_j},
  \end{split}
  \label{eq:cl1phid-2}
\end{align}
and
\begin{equation}
 \begin{split}
  \mathcal{C}_{l_1}(\ket{\varphi^{(d)}_2})=&\abs{\beta_d}^2C_{l_1}(\ket{\varphi^{(d-1)}_2})+2\abs{\alpha_d\beta_d}\sqrt{1+C_{l_1}(\ket{\varphi^{(d-1)}_2}}).
  \label{eq:cl1phi2d}
 \end{split}\end{equation}
 Eqs. (\ref{eq:cl1phi1d}) and (\ref{eq:cl1phi2d}) show that $\mathcal{C}_{l_1}(\ket{\phi^{(d)}_1})<\mathcal{C}_{l_1}(\ket{\phi^{(d)}_2})$ if   $\mathcal{C}_{l_1}(\ket{\varphi^{(d-1)}_1})<\mathcal{C}_{l_1}(\ket{\varphi^{(d-1)}_2})$.

By the definition of the relative entropy of coherence, we have
\begin{align}
  \begin{split}
    \mathcal{C}_r(\ket{\varphi^{(d-1)}_1})=&-\sum_{i=1}^{d-1}\abs{a_i}^2\log_2\abs{a_i}^2,
  \end{split}
  \label{eq:crphid-1}
\end{align}
and
\begin{align}
  \begin{split}
    \mathcal{C}_r({\ket{\varphi^{(d)}_1}})=&-\abs{\beta_d}^2\log_2\abs{\beta_d}^2\\
    &-\sum_{i=1}^{d-1}\abs{\alpha_d}^2\abs{a_i}^2\log_2(\abs{\alpha_d}^2\abs{a_i}^2)\\
    =&-\abs{\beta_d}^2\log_2\abs{\beta_d}^2-\abs{\alpha_d}^2\log_2\abs{\alpha_d}^2\\
    &-\abs{\alpha_d}^2\sum_{i=1}^{d-1}\abs{a_i}^2\log_2\abs{a_i}^2\\
    =&\abs{\alpha_d}^2C_r(\ket{\varphi^{(d-1)}_1})+H(\abs{\alpha_d}^2),
  \end{split}
  \label{eq:crphi1d-1}
\end{align}
where $H(x)=-x\log_2x-(1-x)\log_2(1-x)$ is the binary Shannon entropy function,
and similarly,
\begin{align}
  \begin{split}
    \mathcal{C}_r(\ket{\varphi^{(d-1)}_2})=&-\sum_{i=1}^{d-1}\abs{b_i}^2\log_2\abs{b_i}^2,
  \end{split}
  \label{eq:crphid-3}
\end{align}
and
\begin{equation}
  \mathcal{C}_r({\ket{\varphi^{(d)}_2}})=\abs{\alpha_d}^2C_r(\ket{\varphi^{(d-1)}_2})+H(\abs{\alpha_d}^2).\label{al}
 \end{equation}
Eqs. (\ref{eq:crphid-3}) and (\ref{al}) show that $\mathcal{C}_r(\ket{\varphi^{(d)}_1})>\mathcal{C}_r(\ket{\varphi^{(d)}_2})$ if  $\mathcal{C}_r(\ket{\varphi^{(d-1)}_1})>\mathcal{C}_r(\ket{\varphi^{(d-1)}_2})$. This completes the proof of the theorem. With this theorem, it is easy to obtain the conclusion related to Eq. (\ref{lyt}).

\begin{acknowledgements}
This work was supported by NSF China through Grant No. 11575101 and the National Basic Research Program of China through Grant No. 2015CB921004.
\end{acknowledgements}

\end{document}